\documentstyle[editedvolume,numreferences,epsfig]{crckapb} 

\begin{opening}
\title{QUARK GLUON PLASMA \protect\\
       IN A+A COLLISIONS AT CERN SPS}

\author{Marek Ga\'zdzicki}
\institute{Institut f\"ur Kernphysik, University of Frankfurt\\
           August Euler Str. 6, D-60486 Frankfurt, Germany }

\end{opening}

\runningtitle{QUARK GLUON PLASMA ... }

\begin{document}


\vspace{1cm}
\begin{center}
{\it Invited talk given on `Workshop on Nuclear Matter in
Different Phases and Transitions', March 31 -- April 10, 1998,
Les Houches, France}
\end{center}

\section{Introduction}

At the final state of high energy nuclear collisions many new particles
appear.
They are massive and extended objects: hadrons and hadronic resonances.
What is the nature of particle creation in strong interactions?
How does matter look like in
a state of very high energy density which is created during the collision
of two nuclei?
These questions motivate a broad experimental programme in which
properties of high energy nuclear collisions are investigated
\cite{QM97}.

Due to a lack of a calculable theory of strong interaction
the interpretation of the experimental results has to rely on phenomenological
approaches.
The first proposed models of high energy collision process were
statistical models of the early stage \cite{Fe:50,La:53},
the stage in which the excitation of the
incoming matter takes place.
In their original formulations the models failed to reproduce
experimental results.
However, when a broad set of the  data became available
\cite{Ga:95,Ga:96}, it was realized
\cite{Ga:95a,Ga:97} that after necessary
generalization a statistical approach to the early stage
gives surprising agreement with the  results.
It could be therefore used as a tool to identify the
properties of the state created at the early stage
and answer the question whether this state is in the form of
Quark Gluon Plasma (QGP) -- a quasi ideal gas of deconfined
quarks and gluons.

A special role in this study
is played by the entropy \cite{Va:82}
(at high collision energy carried mainly by
final state pions)  and heavy flavours (strangeness, charm)
 production \cite{Ko:86,Ka:86,Ma:86}.
It can be argued that they  are insensitive to the late stages
of the collision  and therefore
they carry information on  the early stage.

In the first part of this contribution we briefly review
a history of data collection and interpretation of the
results  
on which further presented picture of the early stage
of A+A collisions is based. 
In the second part of the contribution 
basic assumptions and results of
a statistical model
of the early stage of the A+A collisions 
are presented.
Conclusions close the paper.

\section{History}

\vspace{0.2cm}
\noindent
1988: {\it Strangeness enhancement in central S+S collisions at 200 
A$\cdot$GeV.} \\
The NA35 Collaboration presented first results on strangeness
production in central S+S collisions at 200 A GeV \cite{Ga:88}.
It was demonstrated that the production of strange and antistrange
hadrons relative to nonstrange hadrons is increased by a factor of about
2 in S+S collisions in comaprison to nucleon--nucleon (N+N)
interactions at the same energy per nucleon.
This conclusion can be  reached due to large acceptance of
the NA35 streamer chamber  and the measurement of the main carriers
of strangeness produced in the collision process: 
(anti)kaons and $\Lambda$ hyperons.
The observation of strangeness enhancement rules out models
of A+A collisions based on a independent superposition of N+N interactions
\cite{Ga:88}.

\vspace{0.2cm}
\noindent
1992: {\it Absence of strangeness enhancement in p+A interactions at
200 GeV.} \\
The compilation of data on strange and nonstrange hadron production
in p+A interactions 
at 200 GeV leads to the conclusion that the strangeness enhancement
effect is not present in these reactions \cite{Bi:92}.
This suggests that the effect observed in central S+S collisions is not due
to secondary hadronic processes.
During the last 10 years there were numerous attempts to reproduce
strangeness enhancement effect in A+A collisions at SPS 
within string--hadronic models including hadronic rescattering \cite{Od:98}.
No satisfactory description of the data is reached.
This supports preliminary conclusion based on p+A results.

\vspace{0.2cm}
\noindent
1994: {\it Pion suppression in central A+A collisions at low energies.} \\
The first results on pion production in
Au+Au collisions at 11 A$\cdot$GeV (AGS BNL) are
shown \cite{Ha:94}.
The compilation of results on pion production in N+N interactions
and central A+A collisions at AGS and lower energies leads to the 
conclusion \cite{Ga:95} that
the number of produced pions per nucleon participating in the collision
is significantly lower in A+A collisions than in N+N interactions at the
same energy per nucleon.
The observed scaling properties and the magnitude of the pion suppression
effect suggest that it can be caused by
the pion and $\Delta$ absorption \cite{Ga:97a} which takes place during the
expansion of hadronic matter.

\vspace{0.2cm}
\noindent
1994: {\it Pion enhancement in central S+S collisions at 200 A$\cdot$GeV.} \\
The pion suppression effect is not present in central S+S collisions
at 200 A$\cdot$GeV. In contrary, the number of pions per participant is
larger than in the N+N interactions at the same collision energy \cite{Ga:95}.
It is also observed \cite{Ga:95a} 
that the energy dependence of pion (entropy) production in
N+N interactions is consistent with that predicted by the statistical
model of the early stage formulated 
about 50 years ago by Fermi and Landau \cite{Fe:50,La:53}.
Finally it is pointed out that
the transition from the pion suppression at low energy A+A
collisions to pion enhancement observed
 at SPS energy can be due to transition to
deconfined matter which takes place between AGS and SPS energies.
A first version of the statistical model of the early stage is
formulated. The experimental results on pion production
analyzed within this approach indicate that the transition 
leads to the increase of the effective number of degrees of
freedom by a factor of about 3 \cite{Ga:95a}.

\vspace{0.2cm}
\noindent
1995: {\it Pion saturation in central A+A collisions at SPS.} \\
The NA49 experiment presents first results on pion
production in central Pb+Pb collisions at 158 A$\cdot$GeV \cite{Ma:95}.
It is observed that the pion to participant ratio is similar
in central S+S and Pb+Pb collisions.
This result agrees with the prediction of statistical model
of the early stage used for the interpretation of the
S+S results \cite{Ga:95a}.

\vspace{0.2cm}
\noindent
1995: {\it Energy dependence of strangeness production in A+A collisions.} \\
The data on strangeness production in Au+Au collisions at AGS
are obtained \cite{Vi:95}.
These results together with other compiled data on strangeness
production in A+A collisions and N+N interactions lead to the
following observations \cite{Ga:96}.
The strangeness to pion ratio increases by a factor of about 2 between 
AGS and SPS energies for N+N interactions.
The corresponding ratio for central A+A collisions is similar
for AGS and SPS energies.
These experimental findings are interpreted as a result of the
transition to QGP taking place between AGS and SPS energies.
This interpretation, again based on the statistical model, is consistent 
with the interpretation of the pion data.

\vspace{0.2cm}
\noindent
1996: {\it Strangeness saturation in central A+A collisions at SPS.} \\
The NA49 experimant presents first results on strangeness production in
central Pb+Pb collisions at 158 A$\cdot$GeV \cite{Jo:96}.
It is observed that the strangeness to pion ratio for central S+S collisions
and Pb+Pb collisions is similar.
This saturation of strangeness production is 
expected in the statistical approach~\cite{Ga:96}.

\vspace{0.2cm}
\noindent
1997: {\it Quantitative agreement: QGP and A+A results at SPS.
      } \\
It is shown that the pion and strangeness yields
measured for A+A collisions at SPS are in quantitative agreement
with the calculations 
made under assumption that a QGP is created in the early stage of
collisions~\cite{Ga:97}.

\vspace{0.2cm}
\noindent
1996--1998: {\it $J/\Psi$ saturation in Pb+Pb collisions 
at 158 A$\cdot$GeV.} \\
The NA50 experiment presents results on $J/\Psi$ production in
Pb+Pb collisions at 158 A$\cdot$GeV as a function of transverse energy
\cite{Go:96}.
It is shown \cite{Ga:98} that the  $J/\Psi$  to  pion ratio
seems to be independent of the collision centrality.
Thus the dependence of  $J/\Psi$ multiplicity
on the volume of the colliding matter (i.e. $\langle J/\Psi \rangle
\sim V $) is the same as dependence of pion and strangeness yields.
This suggests to consider charm production in A+A collisions
in the statistical model.
It is argued that this dependence is expected in the statistical
model \cite{Ga:98} when the charm production is assumed to be governed
by the phase space, in full analogy to the entropy  and strangeness
treatment.

\section{A Model of the Early Stage of A+A Collisions}

Based on the experimental results and interpretation 
ideas sketched in the previous section
a statistical model of the early stage of A+A collisions
was formulated \cite{Ga:98}. In the following we briefly present 
its main assumptions
and results.  

\subsection{Formulation of the Model}

\vspace{0.1cm}
\noindent
-- The basic assumption of our model is that the 
production of new particles (e.g. quarks and gluons)
in the early stage of nucleus--nucleus collisions is a statistical process.
Thus formation of all microscopic
states allowed by conservation laws
is equally probable.
This means that the probability to produce a given macroscopic
state is proportional to the total number of its microscopic
realizations, i.e.
a macroscopic state 
probability $P$ is
\begin{equation}
P \sim e^S,
\end{equation}
where $S$ is the entropy of the macroscopic state.

\vspace{0.1cm}
\noindent
-- As the particle creation process does not produce net baryonic,
flavour and
electric charges only states  with the total
baryon, flavour and electric
numbers equal to zero should be considered.
Thus the properties of the created state are entirely defined by the
volume in which production takes place, 
the  available energy  and a
partition function.  
In the case of collisions of large nuclei
the  thermodynamical approximation can be used and
 the dependence on the volume and the energy
reduces to the dependence on the energy density.
The state properties can be given in the form of a equation of state.

\vspace{0.1cm}
\noindent
-- We assume that the 
early stage entropy creation takes
place in the volume equal to the volume of the Lorentz contracted
nucleus:
\begin{equation}\label{volume}
V = \frac {V_0} {\gamma},
\end{equation}
where $V_0 = 4/3 \pi r_0^3 A$ and $\gamma = \sqrt{s}_{NN}/(2 m_N)$.
The $r_0$ parameter is taken to be 1.30 fm in order to fit the mean
baryon density in the nucleus, $\rho_0 = 0.11$ fm$^{-3}$.

\vspace{0.1cm}
\noindent
-- Only a fraction of the total 
energy in A+A collision  is transformed
into the energy of created particles.
This is because
a part of the energy is carried net baryon number which
is conserved during the
collision. 
This available energy can be expressed as:
\begin{equation}\label{energyin}
E = \eta (\sqrt{s}_{NN} - m_N)~A.
\end{equation}
The parameter $\eta$ is assumed to be independent of
the collision energy and the system size as suggested by the experimental data.
This experimental observation is usually justified by a model of a
quark--gluon structure
of the nucleon \cite{Po:74}. 
The value of  $\eta$ used for the numerical calculations
is 0.67 \cite{Ba:94}.

\vspace{0.1cm}
\noindent
-- In order to predict a probability of creation of a given 
macroscopic state all possible degrees of freedom
and interaction between them should be given
in the form of the partition function.
In the case of large enough volume the grand canonical approximation 
can be used and the state properties can be given in the form of
an equation of state.  
The question of how one can use this equation of state 
to calculate the space--time evolution (hydrodynamics)
of the created system requires a special separate study. 
 
\vspace{0.1cm}
\noindent
-- The most elementary particles of strong interaction are quarks
and gluons.
In the following we consider $u$, $d$, $s$ and $c$ quarks and the corresponding
antiquarks with the internal number of degrees of freedom equal to
6 (3 colour states $\times$ 2 spin states). In the entropy evaluation the
contribution of $c$, $b$ and $t$ quarks can be neglected due to 
their  large masses. 
The internal number of degrees of freedom for gluons is 16
(8 colour states $\times$ 2 spin states).
The masses of gluons and nonstrange (anti)quarks are taken to be 0,
the strange and charm (anti)quark masses are taken to be 
175 MeV \cite{Le:96} 
and 1.5 GeV, respectively.  
 
\vspace{0.1cm}
\noindent
-- In the case of creation of colour quarks and gluons the
equation of state is assumed to be
the ideal gas equation of state modified by the bag constant $B$ in order
to account for 
 a strong interaction between quarks and gluons
and the surrounding  QCD vaccum (see e.g. \cite{bag}):
\begin{equation}
p = p^{id} - B~,~~~
\varepsilon = \varepsilon^{id} + B,
\end{equation}
where $p$ and
$\varepsilon$ denote pressure and energy density, respectively,
and B is the bag constant selected to be 600 MeV/fm$^3$.
The equilibrium state  
defined above is  called
Quark Gluon Plasma or
Q--state.

\vspace{0.2cm}
\noindent
8. At the final freeze--out stage of the collision 
the degrees of freedom are hadrons~-- 
extended and massive objects composed of (anti)quarks and
gluons.
Due to their finite proper volume hadrons can exist in their well defined
asymptotic states only at rather low energy density.
Estimates $\epsilon < 0.1\div0.4$~GeV/fm$^3$ for the hadron gas with van der
Waals excluded volume have been found in Ref.~\cite{Ye:97}.
 In the early stage of A+A collisions such a low energy density states  
could be created only at very low collision energies of a few GeV per
nucleon.
Asymptotic hadronic states can be questioned as a possible degrees
of freedom in the early stage also on the base of our current
understanding of $e^+ + e^-$ anihilation process, where
the initial degrees of freedom are found to be colourless
$q\overline{q}$ pairs~\cite{Pe:82}.

Guided by this considerations
we assume that at collision energies lower than the energy needed
for a QGP creation
the early stage effective degrees of freedom can be approximated
by 
point--like colourless  bosons.
This state is called W--state (White state).
The nonstrange degrees of freedom which dominate 
the entropy production
are taken to be massless, as seems to be suggested by the
original analysis of the
entropy production in N+N and A+A collisions \cite{Ga:95a}.
Their internal number of degrees of freedom was fitted to the same
data
\cite{Ga:95a,Ga:97} 
to be about 3 times
lower than the
internal number of effective
degrees of
freedom for the QGP
(16 + (7/8)$\cdot$36 $\cong$ 48).
Thus it is taken to be 48/3 = 16.
The mass of strange degrees of freedom is assumed to be 500 MeV, 
equal to the kaon mass.
The internal number of strange degrees of freedom is estimated to
be 14 as extracted from the fit to the strangeness data at AGS.
The phenomenological reduction factor 3 is used 
in our numerical calculations  between the total number of
degrees of freedom for Q--state and nonstrange  W--state
because of the different magnitude of strangeness suppression 
due to different
masses of strangeness carriers in both cases.
The ideal gas equation of state is selected.

\vspace{0.1cm}
\noindent
-- We assume that the only process which changes the entropy content
of the produced matter during the expansion, hadronization and
freeze--out is an equilibration with the baryonic subsystem.
It was argued that it leads to  entropy transfer to baryons
which corresponds to the effective absorption of about 
0.35 $\pi$--mesons per baryon
\cite{Ga:95a,Ga:97a}.
This interaction causes also that the produced hadrons in the final
state do not obey symmetries of the early stage production
process, i.e. there are non-zero baryonic number and electric
charge in the final hadron state.

\vspace{0.1cm}
\noindent
-- It is assumed that the total number of $s$ and $\overline{s}$
quarks is conserved during the expansion, hadronization and 
freeze--out.

\subsection{Main Results}

-- For large enough volume of the system the grand canonical
approximation is applicable.
In the case of the formulated above model this approximation can
be safely used already for central Si+Al collisions at AGS energy
when entropy and strangeness are considered.
It is shown that this approximation is also valid starting from
central S+S collisions at SPS when charm production is calculated.
Thus the first consequence of the model is that entropy, strangeness
and charm yields per volume of the state created  in the early stage
(or equivalently per participating nucleon) should be independent of
the size of the colliding nuclei at SPS energy.
This agrees with the experimental data as shown in Fig. 1. 

-- The model predicts that at low collision energies a pure W--state
is created whereas at high energy a pure Q--state is produced.
There is an intermidiate energy region in which W-- and Q--states
coexist.
For the selected parameters of the model the mixed state starts at
$p_{LAB} $ = 30 A$\cdot$GeV and ends at $p_{LAB} $ = 65 A$\cdot$GeV.
Thus at SPS energies (158--200 A$\cdot$GeV) a pure QGP is predicted
to be created in the early stage.
A quantitative agreement of the model with the results on 
entropy and strangeness in central A+A collisions at SPS is observed
(see solid lines in Fig. 1). 

-- The model predicts that the transition from W--state to Q--state
should be associated with the rapid increase of the pion multiplicity
and an nonmonotonic energy dependence of the strangeness to pion ratio.
The corresponding model results are shown in Fig. 2 together with the 
experimental data.

\section{CONCLUSIONS}

We conclude that a broad set
of experimental data is in agreement with the hypothesis
that QGP is created in central A+A (S+S and Pb+Pb) collisions at the SPS.
Carefull experimental investigation of the A+A collisions in the energy
region between top AGS and SPS energies is needed.

\vspace{1cm}

{\bf Acknowledgements}
I would like to thank to my collaborators M.I.~Gorenstein, St.
Mr\'owczy\'nski and D. R\"ohrich.

\newpage



\newpage

\begin{figure}[p]
\epsfig{file=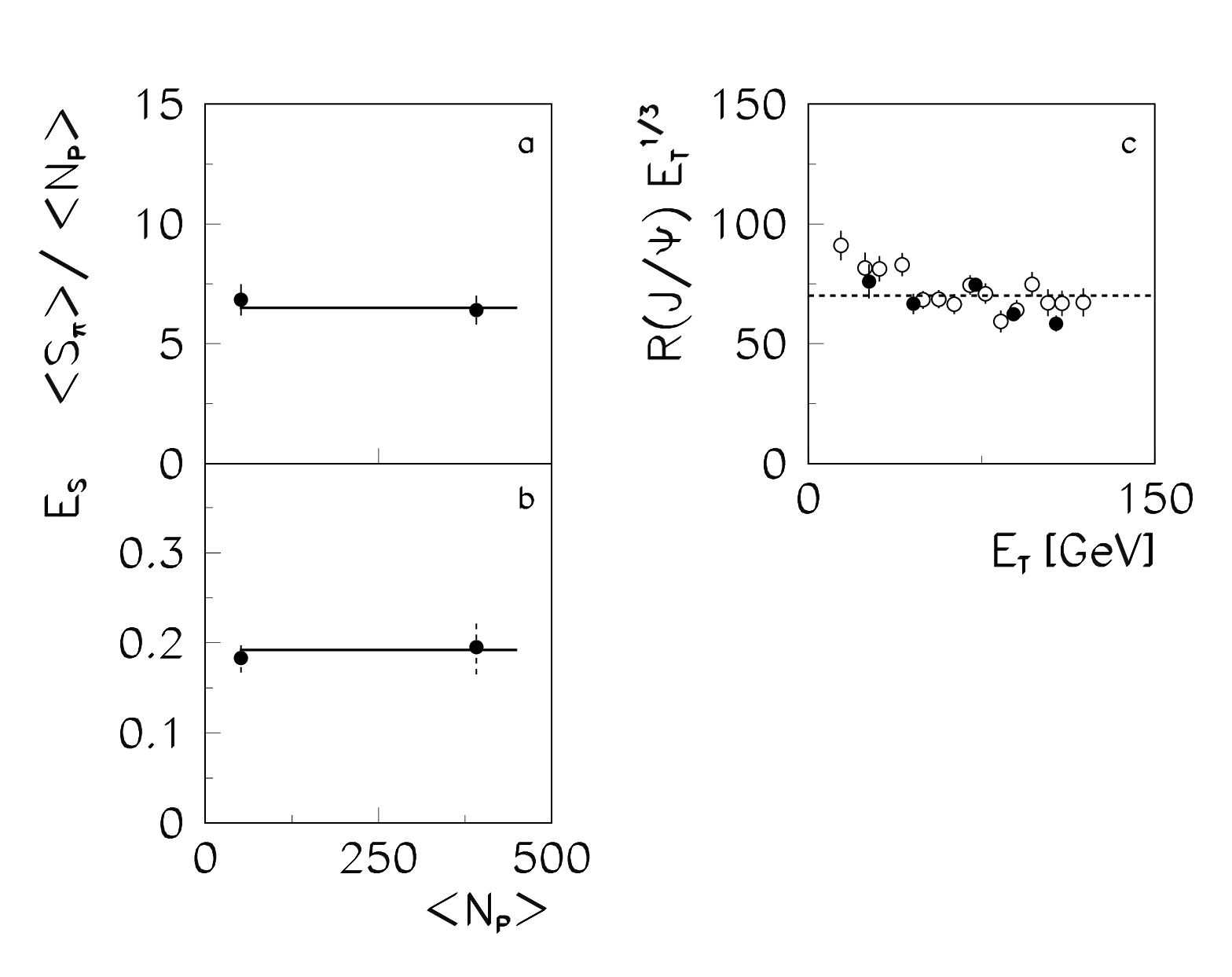,width=14cm}
\caption{
The experimental measures of (quantities proportional to)
entropy to participant (a), strangeness to pion (b) and
$J/\Psi$ to pion (c) ratios as a function of the volume of the colliding
matter for A+A collisions at SPS.
The most right points on the plots correspond to central Pb+Pb
collisions.
The solid lines on Figs. 1a and 1b show results of calculations made within
the statistical model of the early stage.
The resulting magnitude is defined by the properties of the QGP.
The model predicts also indepedence of the charm to pion ratio
of the volume of the colliding matter for large enough systems.
This prediction seems to be in qualitative agreement with the results for 
$J/\Psi$ to pion ratio (Fig. 1c). 
A quantitative comparison requires an additional model
of $J/\Psi$ formation from the hadronizing QGP.
}
\label{fig1}
\end{figure}

\newpage

\begin{figure}[p]
\epsfig{file=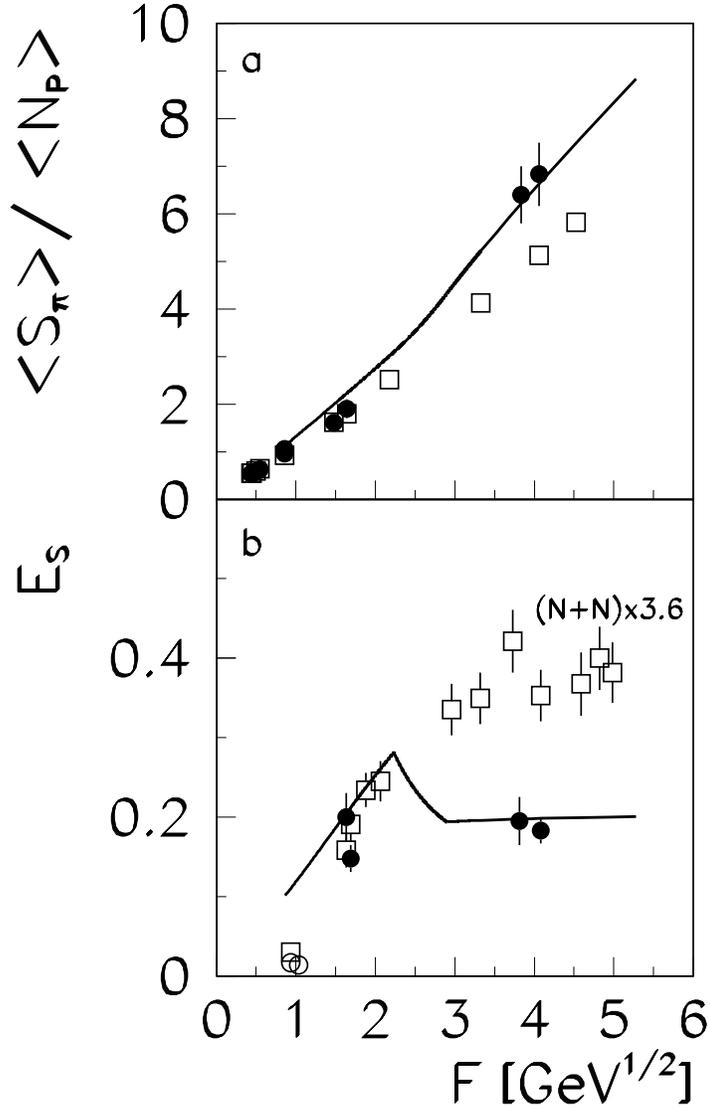,width=11cm}
\caption{
Dependence of the measures of entropy to participant (a) and
strangeness to pion (b) ratios on the collision energy
($F~ =~ (\sqrt{s}_{NN} - 2 m_N)^{3/4} / (\sqrt{s}_{NN})^{1/4} $)
calculated within the statistical model of the early stage
(solid lines).
The compiled data on central A+A collisions are indicated by close
points; the most right points correspond to the collisions at SPS.
The results for N+N interactions are shown by open squares 
for a comparison.
}
\label{fig2}
\end{figure}

\end{document}